\documentclass[12pt]{article}
\input epsf
\def\Journal#1#2#3#4{{#1} {\bf #2}, #3 (#4)}
\newcommand{\NPB}[3]{\emph{ Nucl.~Phys.} \textbf{B#1} (#2) #3}   
\newcommand{\PLB}[3]{\emph{ Phys.~Lett.} \textbf{B#1} (#2) #3}   
\newcommand{\PRD}[3]{\emph{ Phys.~Rev.} \textbf{D#1} (#2) #3}   
\newcommand{\PRL}[3]{\emph{ Phys.~Rev.~Lett.} \textbf{#1} (#2) #3}   
\newcommand{\ZPC}[3]{\emph{ Z.~Phys.} \textbf{C#1} (#2) #3}

\def\dalemb#1#2{{\vbox{\hrule height .#2pt
        \hbox{\vrule width.#2pt height#1pt \kern#1pt
                \vrule width.#2pt}
        \hrule height.#2pt}}}

 \def\bd{\begin{document}} \def\ed{\end{document}}
\def\ds{\documentstyle} \let\fr=\frac \let\bl=\bigl \let\br=\bigr
\let\Br=\Bigr \let\Bl=\Bigl 
\let\bm=\bibitem
\let\na=\nabla
\let\pa=\partial \let\ov=\overline
\def\ie{{\it i.e.\ }} 
\newcommand{\pr}{\paragraph{}}
\newcommand{\be}{\begin{equation}}
\newcommand{\ee}{\end{equation}}
\newcommand{\beba}{\begin{equation}\begin{array}{lcl}}
\newcommand{\eaee}{\end{array}\end{equation}}
\newcommand{\bea}{\begin{eqnarray}}
\newcommand{\eea}{\end{eqnarray}}
\newcommand{\ba}{\begin{array}}
\newcommand{\ea}{\end{array}}
\newcommand{\td}{\tilde}
\newcommand{\norsl}{\normalsize\sl}
\newcommand{\ns}{\normalsize}
\newcommand{\refs}[1]{(\ref{#1})}
\def\simlt{\mathrel{\lower2.5pt\vbox{\lineskip=0pt\baselineskip=0pt
           \hbox{$<$}\hbox{$\sim$}}}}
\def\simgt{\mathrel{\lower2.5pt\vbox{\lineskip=0pt\baselineskip=0pt
           \hbox{$>$}\hbox{$\sim$}}}}

\def\Journal#1#2#3#4{{#1} {\bf #2}, #3 (#4)}

\def\NCA{\em Nuovo Cimento}
\def\NIM{\em Nucl. Instrum. Methods}
\def\NIMA{{\em Nucl. Instrum. Methods} A}
\def\NPB{{\em Nucl. Phys.} B}
\def\PLB{{\em Phys. Lett.}  B}
\def\PRL{\em Phys. Rev. Lett.}
\def\PRD{{\em Phys. Rev.} D}
\def\ZPC{{\em Z. Phys.} C}
\def\st{\scriptstyle}
\def\sst{\scriptscriptstyle}
\def\mco{\multicolumn}
\def\epp{\epsilon^{\prime}}
\def\vep{\varepsilon}
\def\ra{\rightarrow}
\def\ppg{\pi^+\pi^-\gamma}
\def\vp{{\bf p}}
\def\ko{K^0}
\def\kb{\bar{K^0}}
\def\al{\alpha}
\def\ab{\bar{\alpha}}
\def\be{\begin{equation}}
\def\ee{\end{equation}}
\def\bea{\begin{eqnarray}}
\def\eea{\end{eqnarray}}
\def\CPbar{\hbox{{\rm CP}\hskip-1.80em{/}}}
\begin{document}
\thispagestyle{empty}
\rightline{\normalsize\sf hep-ph/0004240}
\rightline{\normalsize CERN-TH/2000-125}
\rightline{\normalsize CPHT-S041.0400}
\rightline{\normalsize April 2000}
\vskip 1.0truecm
\centerline{\Large\bf LIMITS ON THE SIZE OF EXTRA DIMENSIONS\footnote{Based on talks given
by I.A. at Pascos99 \cite{pascos} Lake Tahoe, California, 10-16 December 1999 and by \\ K.B. at Rencontres de Moriond \cite{moriond} on Electroweak Interactions and Unified Theories, Les Arcs, France, March 11-18, 2000.}}
\vskip 1.truecm
\centerline{{\large\bf I. Antoniadis}~$^a$ and  
{\large\bf K. Benakli~$^b$}}
\vskip .5truecm
\centerline{{\it $^a$Centre de Physique Th{\'e}orique, Ecole Polytechnique,
91128 Palaiseau, France}
\footnote{Unit{\'e} mixte du CNRS et de l'EP, UMR 7644.}}
%\centerline{\it Ecole Polytechnique, 91128 Palaiseau, France}
\vskip .5truecm
\centerline{{\it $^b$CERN Theory Division
 CH-1211, Gen{\`e}ve 23, Switzerland}}
\vskip .5truecm

\centerline{\bf\small ABSTRACT}
\vskip .4truecm

We give a brief summary of present bounds on the size of possible 
extra-dimensions  from collider experiments.

\hfill\break
\vfill\eject

\section{Introduction}

In how many dimensions do we live? Could they be more than the four we
 are aware of? If so, why don't we see the other dimensions? Is there
 a way to detect them?. While the possibility of extra-dimensions has
 been considered by physicists  for long time, a compelling reason for
 their existence has arisen with  string theory. It  seems that a
 quantum theory of gravity requires that we live in more than  four
 dimensions, probably in ten or eleven dimensions.  The remaining
 (space-like) six or seven dimensions are hidden to us: observed
 particles do not propagate in them.  The theory  does not tell us yet
 why  four and only four have been  accessible to us. However, it
 predicts that this is only a low-energy effect: with increasing energy,
 particles which propagate in a higher dimensional space could be
 produced. What is the  value of the needed high energy scale? 
 could it be just close by, at reach of near  future experiments?

Another scale which appears in our attempts to answer the previous
questions is related to  the extended  nature of fundamental
objects.  It is the scale at which  internal degrees of freedom are
excited. In string theory this scale $M_s$  is related to the string
tension and sets the mass of the first heavy oscillation mode. The
point-like behavior of known particles as  observed at present
colliders allows to conclude that $M_s$ has to be higher than a
few hundred GeV. However to answer the question of what
energies should be reached before starting to probe this substructure of
the ``fundamental particles'', more precise determination of
experimental lower bounds on $M_s$ and understanding the assumptions
behind them is needed.

It is the aim of this talk as to provide a short summary of the present 
status of limits on these scales of new physics: 
extra-dimensions and string-like sub-structure of matter.

\section{ Hiding Extra-Dimensions }

There is a simple and elegant way to hide the extra-dimensions:
compactification. It is simple because it relies on an elementary
observation. Suppose that  the extra-dimensions form, at each point of
our four-dimensional space, a  $D$-dimensional torus of
volume $(2 \pi)^D  R_1 R_2 \cdots R_D$. 
The $(4+D)$-dimensional Poincare invariance is replaced by a four-dimensional
one times the symmetry group of the  $D$-dimensional
space which contains translations along the $D$ extra directions. The
$(4+D)$-dimensional momentum satisfies the mass-shell condition
$P_{(4+D)}^2 = p_0^2-p_1^2-p_2^2-p_3^2-\sum_i p_i^2 = m_0^2$ and looks from
the four-dimensional point of view as a (squared) mass $ M^2_{KK}=
p_0^2-p_1^2-p_2^2-p_3^2=m_0^2+\sum_i p_i^2$. Assuming periodicity of
the wave functions along each compact direction, one has $p_i = n_i/R_i$
which leads to:
\be
M^2_{KK}\equiv M^2_{\vec n} = m_0^2 +\frac {n_1^2}{R_1^2} +
\frac {n_2^2}{R_2^2}+ \cdots  +\frac {n_D^2}{R_D^2}\, ,
\label{KKdef}
\ee
with $m_0$ the four-dimensional mass and $n_i0$ non-negative
integers.  The states with $\sum_i n_i \neq 0$ are called Kaluza-Klein
(KK) states. It is clear that getting aware of the $i$th
extra-dimension  would require experiments that probe at least an
energy of the order of  $min(1/ R_i)$ with sizable couplings of the
KK  states to four-dimensional matter.

Let us discuss further some properties of the KK states that will be useful
for us below.  We parametrise the ``internal'' $D$-dimensional box by $y_i
\in [-\pi R_i, \pi R_i]$,  $i= 1,\cdots, D$ while  the four-dimensional
Minkowski spacetime is spanned by the coordinates $x^\mu$,   
$\mu= 0, \cdots 3$. It is useful to choose for
the KK wave functions the basis:
\be
\Phi^{\alpha}_{{\vec n},{\vec e}} (x^\mu,y_i)= \Phi^{\alpha} (x^\mu) \,  
\prod_i \left[  
(1-e_i)  \cos (\frac {n_i y_i}{R_i}) + e_i \sin (\frac {n_i y_i}{R_i})
\right]\, ,
\label{KKmodes}
\ee 
where the vector ${\vec n}= (n_1, n_2, \cdots, n_D)$ gives  the
 energy of the state following Eq.~\ref{KKdef} while  $ {\vec e}
 =(e_1,\cdots, e_D)$ with $e_i = 0$  or 1 correspond to a choice of 
cosine or sine dependence in the coordinate $y_i$, respectively. 
 The index ${\alpha}$ refers to other quantum numbers of $\Phi$.

The simplest example of the models we will be using for getting
experimental bounds are  obtained by gauging the  $Z_2$ parity:
$y_i \rightarrow - y_i \,  {\rm mod} \,   2\pi R_i$. This leads
to compactification on segments of size $\pi R_i$. In general, the
consistency of  this ``orbifold'' projection  implies that
the $Z_2$ space parity  should be associated with a $Z_2$  action on 
the internal quantum numbers $\alpha$ of $\Phi$. As a result one has 
the following properties:

\begin{itemize}

\item Only states invariant under this $Z_2$ are kept while the others
are  projected out. There are two  classes of states left in the
theory: those for which $\Phi^{(even)} (x^\mu)$  is even under $Z_2$
action and $e_i =0$ and those for which $\Phi^{(odd)} (x^\mu)$ is odd and
$e_i =1$. It is important to notice that the latter are not present as light
four-dimensional states   i.e. they have $\sum_i n_i \neq 0$ and thus
always correspond to higher KK states.

\item At the boundaries $y_i=0, \pi R$ fixed by the $Z_2$ action, 
new states $\Phi^{(loc)} (x^\mu)$, have to be
included. These ``twisted'' states are localized at the fixed points. They
can not propagate in the extra-dimension and thus have no KK excitations. 

\item The odd bulk states $\Phi^{(odd)} (x^\mu)$ ($e_i =1$) have a wave
function which vanishes  (the $\sin (\frac {n_i y_i}{R_i})$ in
Eq.~\ref{KKdef} ) at the boundaries. Their coupling  to localized states
involves a derivative along $y_i$. For example three boson interactions of
the form $\partial_i \phi^{(odd))} \phi^{(loc)}\phi^{(loc)}$ can be
non-vanishing.

\item The even states, in contrast, can have non-derivative couplings to
localized states. The gauge couplings for instance are given by:
\be
g_n = {\sqrt{2}}\delta^{-{|\vec{\frac{n}{R}}|^2}/{M_s^2}} g
\label{coupling}
\ee
where $\delta >1$ is a model dependent number ($\delta=1/2$ in the case of
$Z_2$). The ${\sqrt{2}}$ comes from the relative normalization of
$\cos(\frac {n_i y_i}{R_i})$ wave function with respect to the zero mode
while the exponential damping is  a result of tree-level string computations
that we do not present here.

\end{itemize}

Use of compactification is an elegant way to hide extra-dimensions
because some of the quantum numbers and  interactions of the
elementary particles could be accounted to by the  topological and
geometrical properties of the internal space. For instance chirality,
number of families in the standard model, gauge and supersymmetry
breaking as well as as some  selection rules in the interactions of
light states  could be reproduced through judicious choice of more
complicated internal spaces.

\section{ Theoretical constraints }

The basic requirement on the theoretical side is that there exist
 theories that  allow the correct magnitude for the strength of the
 gauge and gravitational  couplings for given compactification and
 string  scales just above the present experimental energies. In the
 simplest  string models, the four-dimensional Planck mass can be
 expressed as:
\be
M^2_{pl}  \equiv  f_{pl}  \frac {(M_s^D V_D)} {g_s^p} M_s^2\, ,
\label{newtonH}
\ee
where $V_D$ is the $D$-dimensional internal volume felt by
gravitational interactions, $g_s$ the string coupling and $p$ an
integer.  The four-dimensional gauge coupling can be written as
\be
\frac {1}{g_{YM}^2} \equiv f_{YM} \frac {(M_s^d V_d)}{g_s^q}\, ,
\label{coupH}
\ee
where $V_d$ is the $d$-dimensional internal volume felt by gauge
interactions, and the coefficients $f_{pl}$, $f_{YM}$ have been computed
for known  classical string vacua. In the lowest order approximation,
they are moduli-independent ${\cal O}(1)$ constants.

In the past, weakly coupled heterotic strings were providing the
most promising framework for phenomenological applications. In this
case, the standard model was considered as descending from the
ten-dimensional $E_8$ gauge symmetry, and we have $V_d =V_D$, $D=d=6$ and
$p=q=2$. Taking the ratio of the two equations, one finds $ \frac
{M_s^2} {M^2_{pl}}= \frac { f_{YM}}{f_{pl}} g_{YM}^2   \sim g_{YM}^2$.
Requiring $g_{YM} \sim {\cal O}(1)$, it was concluded that both the 
string scale $M_s$
and the compactification scale $R^{-1} \equiv V_6^{-1/6}$ had to lie
just below the Planck scale, at energies $\sim 10^{18}$GeV far out of
reach of any near future experiment~\cite{Ant,add}.

The situation changed during recent years~\cite{str} when it was discovered
that string theory provides classical solutions (vacua) where gauge degrees
of freedom live on subspaces i.e. $d < D$ along with the possibility of 
$p \neq q$. For instance, while $D=6$ and $p=2$, $(d,q)=(d,1)$ in type I
and $(d,q)=(2,0)$ in type II or weakly coupled heterotic strings with small
instantons. In these cases, it is an easy exercise to check that both the
string and compactification scales can be made arbitrarily low~\cite{add,scale}.

Lowering  the string scale, one increases the strength of higher
(non-renormalizable) operators leading to the possibility of inducing
exotic processes at experimentally excluded rates. Although an explicit 
string realization of the scenario is necessary in order to have a
satisfactory solution, at the effective field theory level many discrete or
global symmetries can be displayed that forbid these operators.

\section{ Experimental constraints }\label{sec:exp}

\subsection{The scenario:}

In order to pursue further, we need to provide the quantum numbers and 
couplings of the relevant light states. In the scenario we consider:
\begin{itemize}

\item Gravitons~\footnote{ Along with gravitons, string models predict the
presence of other very weakly coupled states as gravitinos, dilatons,
moduli, Ramond-Ramond fields....These might alter the bounds obtained in 
Section~\ref{subsec:miss}.} which describe fluctuations of the metric 
propagate in the whole 10- or 11-dimensional space. 

\item In all generality, the gauge bosons propagate on a $(3+d)$-brane, with 
$d=0,...,6$. However, as we have seen in the previous section, a freedom of 
choice for the values of the string and compactification scales requires
that gravity and gauge degrees of freedom live in spaces with different
dimensionalities. This means that $d_{max} =5$ or 6 for 10- or 11-dimensional
theories, respectively. The value of $d$ represents the number of dimensions
felt by KK excitations of gauge bosons. 

\item The matter fermions, quarks and leptons, are localized on a
3-brane and have no KK excitations.  
Their coupling to KK modes of gauge bosons are given in Eq.~\ref{coupling}.
This is the main assumption in our analysis and limits derived in the next
subsection depend on it. In a more general study it could be relaxed by
assuming that only part of the fermions are localized. However, if all
states are propagating in the bulk, then the KK excitations are stable and a
discussion of the cosmology will be necessary in order to explain why they
have not been seen as isotopes.

\end{itemize}

The possible localization of the Higgs scalar, as well as the possible
existence of supersymmetric partners do not lead to important modifications 
for most of the obtained bounds.

\subsection{Extra-dimensions along the world brane: KK excitations of gauge
bosons}

To simplify the discussion, let us first consider the case $d=1$ where some
of the gauge fields arise from a 4-brane. Since the  couplings  of the
corresponding gauge groups are reduced by  the size of the large dimension
$R_\parallel M_s$ compared to the others, if  $SU(3)$ has KK modes all three
group factors must have. Otherwise it is difficult to reconcile the
suppression of the strong coupling at the string scale with the observed
reverse situation. As a result, there are 5 distinct cases that we denote
$(l,l,l)$, $(t,l,l)$, $(t,l,t)$, $(t,t,l)$ and $(t,t,t)$, where the three
positions in the brackets correspond to the 3 gauge group factors of the
standard model $SU(3)_c\times SU(2)_w\times U(1)_Y$ and those with $l$ feel
the extra-dimension, while those with $t$ (transverse) do not.

%%%%%%%%%%%%%%%%%%%%%%%%%%%%%%%%%%%%%%%%%%%%%%%%%%%%%%%%%%%%%%%%%%%%%
\begin{table}[t]
\caption{Limits on $R^{-1}_\parallel$ in TeV at present and future
colliders. The luminosity  is given in  fb$^{-1}$.\label{tab:para}}
\vspace{0.4cm}
\begin{center}
\begin{tabular}{  | c | c | c | c | l |} 
\hline
  & & & &
\\   Collider & Luminosity &  Gluons  & $W^{\pm}$ & $\gamma + Z$   \\ 
\hline\hline
\mco{5}{|c|}{Discovery of Resonances}   \\ \hline
  LHC     & 100        &  5  & 6  & 6               \\ \hline\hline
\mco{5}{|c|}{Observation of Deviation}   \\ \hline
 LEP 200    &$4\times 200$  & - & -  & 1.9 \\ \hline
 TevatronI & $0.11$ &  -  & - & 0.9  \\ \hline 
 TevatronII & 2 &  -  & - & $1.2$ \\ \hline 
  TevatronII  & 20 &  4 & - & 1.3 \\ \hline 
 LHC & 10& 15   & 8.2  & 6.7 \\ \hline
 LHC & 100& 20   & 14 &  12  \\ \hline
  NLC500 & 75& - & - & 8  \\ \hline
  NLC1000 & 200& - & - & 13  \\ \hline\hline
\end{tabular}
\end{center}
\end{table}
%%%%%%%%%%%%%%%%%%%%%%%%%%%%%%%%%%%%%%%%%%%%%%%%%%%%%%%%%%%%%%%%%%%%%%%%

The experimental signatures of extra-dimensions are of two types:
\begin{itemize}

\item Observation of resonances due to KK excitations. This needs a collider
energy $\sqrt{s} \simgt 1/R_\parallel$ at LHC. The discovery limits in the
case of one extra-dimension are given in table 1.

\item Virtual exchange of the KK excitations which lead to measurable
deviations in cross-sections compared to the standard model prediction. The
exchange of KK states gives rise to an effective operator:
\be
 {\bar \psi_1} {\psi_2} {\bar \psi_3} {\psi_4} 
\sum_{|\vec{n}|} \frac {g^2{(|\vec{n}|)}}{m_0^2 +
\frac{|\vec{n}|^2}{R_\parallel^2}}\, .
\ee
The usual approximation of taking $g^2{(|\vec{n}|)}$ independent of
$|\vec{n}|$ fails for more than one dimension because the sum $\sum_{n_i}
\frac {1}{ n_1^2 +n_2^2 +...}$ becomes divergent. This divergence is
regularized by the exponential damping of Eq.~\ref{coupling}. For
$d>1$ the result depends then on both parameters $R_\parallel$ and
$M_s$. Example of analysis for $d=2$ can be found in Ref.~\cite{ABQ}. The
simpler case of $d=1$ has been studied in detail. Possible reaches
of colliders experiments~\cite{AAB,ABQ} are summarized in table 1.

The effects of exchange of virtual KK modes 
are also constrained by  high precision data~\cite{scc,Delgado}, such
as the fit of the measured values of $M_W$, $\Gamma_{ll}$ and
$\Gamma_{had}$. If the Higgs is assumed to be a bulk state like the gauge
bosons, then one finds $R^{-1} \simgt 3.5$ TeV. Inclusion of $Q_W$
measurement, which does not give a good agreement with the  standard model
itself, raises the bound to $R^{-1} \simgt 3.9$ TeV~\cite{Delgado}. The
presence of a localized Higgs  allows tree-level mixing of different KK, 
 and makes the bounds model dependent~\cite{Delgado}.

\end{itemize}

There are some ways to distinguish the corresponding signals from other
possible origin of new physics, such as models with new gauge bosons. In the
case of observation of resonances, one expects three resonances in  the
$(l,l,l)$ case and two in the  $(t,l,l)$ and $(t,l,t)$ cases, located
practically at the same mass value. This property is not shared by most of
other new gauge boson models. Moreover, the heights and widths of the
resonances are directly related to those of standard model gauge bosons in
the corresponding channels.  In the case of virtual effects, these are not
reproduced by a tail of Bright-Wigner shape and a deep is expected just
before the resonance of the photon+$Z$, due to the interference between the
two. However, good statistics will be necessary.

\subsection{Extra-dimensions transverse to the brane world: KK excitations
of gravitons}\label{subsec:miss}

The localization of (infinitely massive) branes in the $(D-d)$ dimensions 
breaks translation invariance  along these directions. Thus, the
corresponding momenta are not conserved: particles, as gravitons, could be
absorbed or emitted from the brane into the $(D-d)$ dimensions. Non
observation of the effects of such processes  allow us to get bounds on the
size of these transverse extra dimensions. In order  to simplify the
analysis, it is usually assumed that among the $D-d$ dimensions $n$ have
very large common radius $R_{\perp } \gg M_s^{-1}$, while the remaining
$D-d-n$ have sizes of the order of the string length.

During a collision of center of mass energy $\sqrt{s}$, there are 
$(\sqrt{s}R_{\perp})^n$ KK excitations of gravitons with mass
$m_{KK\perp}<\sqrt{s}< M_s$, which can be emitted. Each of these states 
looks from the four-dimensional point of view as a massive, quasi-stable, 
extremely weakly coupled ($s/M^2_{pl}$ suppressed) particle that escapes
from the detector. The total effect is a missing-energy cross section
roughly of order: 
\be
\frac {(\sqrt{s}R_{\perp })^n} {M^2_{pl}} \sim \frac{1}{s} 
{(\frac{\sqrt{s}}{M_s})^{n+2}}
\label{miss1}
\ee
Explicit computation of these effects leads to the bounds given in table 2~\cite{missing}.
The results require some remarks:
\begin{itemize}

\item The amplitude for emission of each of the KK gravitons is taken to be
well approximated by the tree-level coupling of the massless graviton as
derived from General Relativity. Eq.~\ref{coupling} suggests that this is
likely to be a good approximation for $R_{\perp}M_s\gg 1$.

\item The cross-section depends on the size $R_\perp$ of the transverse
dimensions and allows to derive bounds on this {\it physical} scale. 
As it can be seen from Eq.~\ref{coupling}, transforming these bounds to 
limits on $M_s$ there is an ambiguity on different factors involved, such as
the string  coupling. This is sometimes absorbed in the so called
``fundamental quantum gravity scale $M_{(4+n)}$''.  Generically $M_{(4+n)}$
is bigger than $M_s$, and in some cases, as in type II strings or in
heterotic strings with small instantons, it can be many orders of magnitude
higher than $M_s$, so it does not correspond to a scale  where some physical
phenomena open up.

\item There is a particular energy and angular distribution  of the
produced gravitons that arises from the distribution in mass  of KK
states given in  Eq.~\ref{KKdef}. It might be a smoking gun for the
extra-dimensional nature of such observable signal.

\item For given value of $M_s$, the cross section for graviton emission
decreases with the number of large transverse dimensions. The effects
are more likely to be observed for the lowest values of $M_s$ and $n$.

\item Finally, while the obtained bounds for $R_\perp^{-1}$ are 
smaller than those that could be checked in table-top experiments probing 
macroscopic gravity at small distances, one should keep in mind that 
larger radii are allowed if one relaxes the assumption of isotropy, 
by taking for instance two large dimensions with different radii.

\end{itemize}

In table 2, we have also included astrophysical and cosmological
bounds. Astrophysical bounds~\cite{astcos,supernovae} arise from the
requirement that the radiation of gravitons should not carry on too much
of the gravitational binding energy released during core collapse of
supernovae. In fact, the measurements of Kamiokande and IMB for SN1987A 
suggest that the main channel is neutrino fluxes.

The best cosmological bound~\cite{COMPTEL} is obtained from requiring that
decay of bulk gravitons to photons do not generate a spike in the energy
spectrum of the photon background measured by the COMPTEL instrument. The
bulk gravitons  are themselves expected  to be produced just before
nucleosynthesis due to thermal radiation from the brane. The limits assume
that the temperature was at most 1 MeV as nucleosynthesis begins, and become
stronger if the temperature is increased.

%%%%%%%%%%%%%%%%%%%%%%%%%%%%%%%%%%%%%%%%%%%%%%%%%%%%%%%%%%%%%%%%%%%%%
\begin{table}[t]
\caption{Limits on $R_\perp$ in mm from missing-energy
processes.\label{tab:exp3}}
\vspace{0.4cm}
\begin{center}
\begin{tabular}{  | c | c | c | l |} 
\hline
  & & &
\\   Experiment & $R_\perp (n=2)$ & $R_\perp (n=4)$ & $R_\perp (n=6)$ \\ 
\hline\hline
\mco{4}{|c|}{Collider bounds}   \\ \hline

 LEP 2   & $4.8\times 10^{-1}$ & $1.9\times 10^{-8}$  & 
                              $6.8 \times 10^{-11}$ \\ \hline
  Tevatron  &   $5.5 \times 10^{-1}$  & $1.4 \times 10^{-8}$ 
              & $4.1 \times 10^{-11}$ \\ \hline 
  LHC &  $4.5 \times 10^{-3}$   & $5.6\times 10^{-10}$  & 
                              $2.7 \times 10^{-12}$  \\ \hline
  NLC & $1.2\times 10^{-2}$  & $1.2\times 10^{-9}$  & 
                              $6.5 \times 10^{-12}$  \\ \hline\hline
\mco{4}{|c|}{Present non-collider bounds}   \\ \hline
  
SN1987A   &  $3 \times 10^{-4}$   & 
           $1 \times 10^{-8}$ 
                 & $6 \times 10^{-10} $ \\ \hline
COMPTEL &  $5 \times 10^{-5}$   & - & 
                              - \\ \hline
\end{tabular}
\end{center}
\end{table}
%%%%%%%%%%%%%%%%%%%%%%%%%%%%%%%%%%%%%%%%%%%%%%%%%%%%%%%%%%%%%%%%%%%%%%%%

\subsection{Dimension-Eight Operators and Limits on The String Scale:}

At low energies, the interaction of light (string) states is
described by an effective field theory. Non-renormalizable dimension-six
operators are due to the exchange of KK excitations of gauge bosons between
localized states. If these are absent, then there are deviations to the
standard model expectations from dimension-eight operators. There are two
generic sources for such operators: exchange of virtual KK excitations of
bulk fields (gravitons,...) and form factors due to the extended nature of
strings.

The exchange of virtual KK excitations of bulk gravitons is described
in the effective  field theory by an amplitude involving the sum
$\frac {1}{M_p^2}\sum_n \frac {1}{s-\frac{{\vec n}^2}{R_\perp^2}}$. For
$n > 1$, this sum diverges and one cannot compute it in field theory but {it
only} in a fundamental (string) theory. In analogy with  the case of
exchange of gauge bosons, one expects the string scale to act as a cut-off
with a result: 
\be  
A g_s^2 \frac {T_{\mu \nu}T^{\mu \nu} - \frac {1}{1+d_\perp} 
T_\mu^\mu  T_\nu^\nu} {M_s^4}\, .
\label{QFT}
\ee 
The approximation $A = \log{ \frac{M_s^2}{s}}$ for $d_\perp =2$ and 
$A = \frac{2}{d_\perp-2}$ for $d_\perp > 2$ is usually used for 
quantitative discussions. There are some reasons which might invalidate 
this approximation for particular cases. In fact, the result is 
very much model dependent: in type I string models it reflects the 
ultraviolet behavior of open string one-loop diagrams which are model 
(compactification) dependent.

In order to understand better this issue, it is important to remind that in
type I string models, gravitons and other bulk particles correspond to
excitations of closed strings. Their tree-level exchange is described  by a
 cylinder joining the initial $|B in>$ and final $|B out>$ closed
 strings lying on the brane. This cylinder can be be seen on the other
hand as an open string with one of its end-points describing the
 closed (loop)  string $|B in>$, while the other end draws $|B out>$. In
other words, the cylinder can be seen as an annulus which is a one-loop
 diagram of open strings with boundaries $|B in>$ and $|B out>$. Note that
the validity  of this duality, which only assumes the branes to be
Dirichelet branes of  string theory, seems to require the presence of other
weakly interacting closed strings besides gravitons.

More important is that when the gauge degrees of freedom
arise from Dirichelet branes, it is expected that the dominant source
of dimension-eight operators is not the exchange of KK states but
instead the effects of massive open string oscillators~\cite{AAB,string}. These give rise to
contributions to tree-level scattering that behave as $g_s s/M_s^4$.
Thus, they are enhanced by a string-loop factor $g_s^{-1}$ compared to
the field theory estimate based on KK graviton exchanges. Although the
precise value of $g_s$ requires a detail analysis of threshold
corrections, a rough estimate can be obtained by taking 
$g_s\simeq\alpha\sim 1/25$, implying an enhancement by one order of
magnitude. 

What is the simplest thing one could do in practice?.  There are some
processes for which there is only one allowed dimension-eight operator; an
example is $f{\bar f}\rightarrow \gamma\gamma$. The coefficient of this
operator can then be computed in terms of $g_s$ and $M_s$.
As a result, in the only framework where computation of such operators is
possible to carry out, one cannot rely on the effects of exchange of KK
graviton excitations in order to derive bounds on extra-dimensions or the
string scale. Instead, one can use the dimension-eight operator arising
from stringy form-factors.

\section*{Acknowledgments}
This work was partly supported by the EU under TMR contract ERBFMRX-CT96-0090.
We would like to thank E. Accomando, Y. Oz and M. Quir\'os for enjoyable
collaborations, and the organizers of PASCOS and Moriond conference for
invitation and financial support.

\end{document}